# Extreme Ultraviolet Transient Grating Spectroscopy of Vanadium Dioxide

Emily Sistrunk,[1,†] Jakob Grilj,[1,2,†] Jaewoo Jeong,[3] Mahesh G. Samant,[3] Alexander X. Gray,[4,5] Hermann A. Dürr,[4] Stuart S. P. Parkin,[3] and Markus Gühr[1,*]

[1]Stanford PULSE Institute, SLAC National Accelerator Laboratory, 2575 Sand Hill Rd., Menlo Park, CA 94025
[2]Laboratory of Ultrafast Spectroscopy, Ecole Polytechnique Federal de Lausanne, CH 1015 Switzerland
[3]IBM Almaden Research Center, 650 Harry Road, San Jose, CA 95120
[4]Stanford Institute for Materials and Energy Sciences, SLAC National Accelerator Laboratory, 2575 Sand Hill Rd., Menlo Park, CA 94025
[5]Department of Physics, Temple University, Philadelphia, PA 19122
*Corresponding author: mguehr@stanford.edu
[†]Contributed equally.

Nonlinear spectroscopy in the extreme ultraviolet (EUV) and soft x-ray spectral range offers the opportunity for element selective probing of ultrafast dynamics using core-valence transitions (Mukamel *et al.*, Acc. Chem. Res. **42**, 553 (2009)). We demonstrate a step on this path showing core-valence sensitivity in transient grating spectroscopy with EUV probing. We study the optically induced insulator-to-metal transition (IMT) of a $VO_2$ film with EUV diffraction from the optically excited sample. The $VO_2$ exhibits a change in the 3p-3d resonance of V accompanied by an acoustic response. Due to the broadband probing we are able to separate the two features.

Introduction

Ultrashort light pulses in the extreme ultraviolet (EUV) provide opportunities for element selective probing of ultrafast valence dynamics [1]–[4] and are even suggested for element selective multidimensional spectroscopy [5], [6]. Since core and deep valence binding energies differ strongly by element, the core to outer valence transitions are unique for different elements within one compound [7]. Prominent examples for these transitions in the EUV are the 3p-3d absorption features near the M-edge of 3d transition metals [8].

Recent time-resolved transmission spectroscopy in hematite with a broadband EUV probe continuum situated around the M-edge of iron resulted in novel insight on the charge transfer processes in metal oxides [3]. This method is sensitive to excitation induced changes of the imaginary part of the refractive index. To measure this one has to distinguish small signal changes induced by the excitation on top of a large background. Depending on the light source noise characteristics, the background can impose a problem for obtaining sufficient signal to noise. An alternative way of measuring the sample response is based on diffracting the EUV probe from an excitation *grating*. The resulting so called transient grating (TG) spectroscopy, a special variant of the four-wave-mixing method, has been highly successful in IR and visible ultrafast spectroscopy [9]–[12]. The spatial excitation modulation can be achieved by two excitation pulses that interfere on the sample under a small angle. The resulting intensity grating is transformed into an index of refraction grating by the sample. The probe pulse is then scattered off this grating, which means that the detector is free of signal without excitation pulses. TG has been utilized to examine the electronic response of complex materials, electron-phonon coupling, and molecular dynamics [12]–[15]. It has proven a powerful technique in chemistry and materials science, particularly in the infrared, visible and near-ultraviolet [16].

Recently the TG technique has been extended into the EUV range. The EUV probe was reflected from an excited sample to study the acoustic response of a thin metal film [17]. In this case, the grating was not present in the form of a change in refractive index but as a change in the surface height. Using a small spectral band in the EUV range, it was found that the EUV, due to its short wavelength, is extremely sensitive to surface changes, which was further applied in refs [17]–[19]. In this paper, we extend the technique. We operate with a broad spectral bandwidth of 60% which we also disperse by the transient grating. This allows us to study the refractive index response, opening the field of element sensitive probing by EUV-TG experiments. Due to the large bandwidth we can distinguish the index of refraction response from the surface (acoustic) changes which are also present in the sample.

We demonstrate this technique for the example of vanadium dioxide ($VO_2$), which is a strongly-correlated material of particular interest because of a near-room-temperature thermally activated insulator-to-metal transition (IMT) [20]. The IMT can also be induced by light on an ultrafast timescale, important for ultrafast electronics applications (see for instance [21]–[23]). The extensive literature documents a debate over the ultrafast mechanism of the IMT in $VO_2$ [24]–[27]. The use of EUV TG spectroscopy allows for element selective probing of ultrafast photo-induced transitions with a table top laser source.

This paper will provide a description of the transient grating beam line and typical detected signal in the experiment. We will demonstrate the time-resolved spectral response of the $VO_2$ and analyze its features. We will discuss the resolution limits of the experiment and provide outlook for future improvements.

Experiment

$VO_2$ films of 100nm thickness were deposited on a 0.5mm $Al_2O_3$ ($10\bar{1}0$) substrate by pulsed laser deposition in oxygen pressure of 10mTorr and at growth temperature of 700°C [28]. The samples were mounted in a vacuum chamber on a heated holder. The transient grating is created on the surface of the $VO_2$ sample by two crossing pulses of 30fs duration and of 800nm central wavelength (see Fig. 1). The grating period of ~15μm is achieved by angular separation of 4 degrees and an average fluence of 15mJ/cm$^2$ is created on the sample. With 30fs pulses we observe immediate sample damage for average fluence above 30mJ/cm$^2$ and slowly accumulated damage above 20mJ/cm$^2$. The broadband EUV probe consists of high harmonics of 800nm, generated with a loose focusing geometry (~1.5m focal length) in an argon gas cell. It is separated from the IR fundamental by a silicon wafer near Brewster angle and an aluminum filter of 150nm thickness

[29]. All three IR beams are derived from the same 800nm, 30fs, Ti:Sapphire laser. The high harmonics are refocused with a toroidal mirror, and hit the $VO_2$ sample surface at a grazing angle of 22 degrees before reaching the focus at the detector. Harmonics of order 15-27 (which corresponds to photon energies of 23-42eV) are resolved in the first order diffracted signal and integrated using an EUV-sensitive back-thinned CCD camera.

Figure 2a shows the first order diffraction of the EUV spectrum from the transient grating. The CCD was exposed for 2.5s. Spatial integration over a range of 10 pixels leads to the spectrum in Fig 2b. The spectral calibration is performed by calculating the dispersion angle expected from the grating line spacing, which is measured independently by replacing the sample with a BBO crystal and imaging the grating pattern onto a CCD camera. Figure 2c shows the vanadium M-edge via the experimentally measured reflectivity of $VO_2$ in its insulating phase [30] and the calculated M-edge absorption [31] of vanadium.

The delay between the grating excitation and the broadband EUV probe pulses is controlled by a translation stage using a minimum step size of 25fs. Transient grating signals are acquired for 2.5s per data point and summed over more than 5 time delay sweeps. The final time-delayed data scans are averaged over three delay steps to further reduce noise. The EUV pulse front sweeps with a 22 degree grazing angle across a sample region that is simultaneously pumped by the excitation pulses, which leads to a degradation of the time resolution to about 3ps.

The sample temperature is varied from 20 to 75°C to thermally induce the IMT occurring at 70° for this sample. Sample temperature stability is verified by a thermistor pressed onto the sample surface.

Results

The time dependence of the transient response of $VO_2$ was studied with the sample below and above the thermally induced IMT. Time-dependent traces are included in Figure 3 for integrated spectra at harmonic orders (HH) 21-25, corresponding to photon energies of 32 (black), 35 (red), and 39eV (blue). Those three harmonics provide the best signal to noise ratio. On long time-scales the $VO_2$ shows a rise-time of approximately 20ps for both room temperature (in Fig. 3a) and 75°C (in Fig. 3c) samples. The diffracted signal then decreases slowly with oscillations 10's of ps in period. This long-term response is similar for all photon energies, although the decay of the diffracted signal is much faster in the room temperature sample, which decays to roughly 60% by a delay of 150ps. In contrast the 75°C signal experiences an initial decay followed by a plateau at 80% of the maximum signal.

Examining the short term behavior, we see a marked difference in the room temperature (Fig. 3b) and 75°C (Fig. 3d) samples. For the heated phase, which is already metallic, the ultrashort pulses cannot induce the IMT. In this phase, the hot sample behaves similarly at all three photon energies. The cold sample however displays a fast-rising feature only in harmonics with photon energy above 38eV. This signal increases within 3ps, much faster than the other harmonics, and then continues increasing with the same dependence as the other colors.

Discussion

We first discuss the long term response of the sample. As demonstrated in Figure 3a and c the long-term behavior of the transients shows a maximum in diffraction at approximately 20ps. This is followed by a slow decay for cold samples or constant amplitude for hot samples, convoluted with picosecond oscillations. *All photon energies* are modulated in a similar way. This is consistent with behavior reported from multiply reflected acoustic waves in thin metal films [12], [17], and results from surface height changes. In agreement with ref. [17], we find that the EUV diffraction is highly sensitive to thermal expansion of the sample. For the hot sample this is the result of heat deposited by the laser, while in the cold sample the mechanism includes both heating and lattice distortion due to the IMT. We estimate the expansion of the metallic sample due to absorbed laser energy to be on the order of 7Å based on the $\alpha_{33}$ thermal expansion coefficient [32] and reported values for the specific heat capacity [33] and density [34] of $VO_2$. This surface height change results in a sinusoidal phase grating, with calculated diffraction efficiency of 1% averaged over the harmonic spectral bandwidth we collect [9], [13], [35]. From the experiment we estimate a few percent diffraction efficiency, reasonably close to the calculation and in agreement with other reported diffraction efficiencies in the EUV [17]. In the short time behavior of the cold sample (in Fig. 3b) we notice a fast response of *only the highest energy photons* absent in the hot sample. This fast response cannot be merely the result of a surface height change because it does not affect the diffraction at the other photon energies. Therefore it must be related to a change in the index of refraction, $n(\omega)$, as a function of frequency. This demonstrates how the dispersion of a broad spectral range is crucial for separating surface change and index change.

The fast feature recorded in the rising edge of HH25 shows that the excitation prepares an ultrafast response that can only be detected by this harmonic. This is indicative of a shift of the vanadium M-edge (3p-3d features) to lower energy or of new 3p-3d features created around HH25 as a result of the IMT. The IMT must be the cause because the hot, metallic sample does not show this fast response. In contrast to results obtained with monochromatic probing [17], the dispersive sensitivity of the TG geometry allows us to discern spectral features. Whether the energy shift of the absorption feature is due to electron correlation or changes in lattice structure is still unclear. Increased time resolution and better coverage of the entire M-edge are needed to clarify the mechanism.

Conclusions

The first EUV-TG spectrally resolved time-dependent study of VO$_2$ demonstrates the powerful capabilities of our method. Due to the broad bandwidth and the intrinsic dispersion of the transient grating, we succeed in separating an index of refraction change at the M edge of vanadium from the acoustic response of the sample. This change is triggered by the optically induced IMT. We note that accumulated damage to the surface of the sample can create a permanent grating, producing a diffracted signal under the transient signal. This opens the interesting opportunity for future heterodyne measurements [14] allowing for signal amplification and decomposition into real and imaginary material constants.

J.G. and E.S. contributed equally to the experimental set-up, data acquisition, and analysis. J.G. acknowledges support by the European Research Agency via the FP-7 PEOPLE Program (Marie Curie Action 298210). M.G. acknowledges funding via the Office of Science Early Career Research Program through the Office of Basic Energy Sciences, U.S. Department of Energy. This work was supported by the AMOS program within the Chemical Sciences, Geosciences, and Biosciences Division of the Office of Basic Energy Sciences, Office of Science, U.S. Department of Energy.

**References.**
1. C. La-O-Vorakiat, M. Siemens, M. M. Murnane, H. C. Kapteyn, S. Mathias, M. Aeschlimann, P. Grychtol, R. Adam, C. M. Schneider, J. M. Shaw, H. Nembach, and T. J. Silva, "Ultrafast Demagnetization Dynamics at the M Edges of Magnetic Elements Observed Using a Tabletop High-Harmonic Soft X-Ray Source," Phys. Rev. Lett. **103**, 257402 (2009).
2. A. Melnikov, H. Prima-Garcia, M. Lisowski, T. Gießel, R. Weber, R. Schmidt, C. Gahl, N. M. Bulgakova, U. Bovensiepen, and M. Weinelt, "Nonequilibrium Magnetization Dynamics of Gadolinium Studied by Magnetic Linear Dichroism in Time-Resolved 4f Core-Level Photoemission," Phys. Rev. Lett. **100**, 107202 (2008).
3. J. Vura-Weis, C.-M. Jiang, C. Liu, H. Gao, J. M. Lucas, F. M. F. de Groot, P. Yang, A. P. Alivisatos, and S. R. Leone, "Femtosecond M2,3-Edge Spectroscopy of Transition-Metal Oxides: Photoinduced Oxidation State Change in α-Fe2O3," J. Phys. Chem. Lett. **4**, 3667 (2013).
4. Z.-H. Loh and S. R. Leone, "Capturing Ultrafast Quantum Dynamics with Femtosecond and Attosecond X-ray Core-Level Absorption Spectroscopy," J. Phys. Chem. Lett. **4**, 292 (2013).
5. F. Bencivenga, S. Baroni, C. Carbone, M. Chergui, M. B. Danailov, G. D. Ninno, M. Kiskinova, L. Raimondi, C. Svetina, and C. Masciovecchio, "Nanoscale dynamics by short-wavelength four wave mixing experiments," New J. Phys. **15**, 123023 (2013).
6. S. Mukamel, D. Abramavicius, L. Yang, W. Zhuang, I. V. Schweigert, and D. V. Voronine, "Coherent Multidimensional Optical Probes for Electron Correlations and Exciton Dynamics: From NMR to X-rays," Acc. Chem. Res. **42**, 553 (2009).
7. J. Stöhr, *NEXAFS Spectroscopy*. (Berlin; New York: Springer, 1996).
8. M. Martins, K. Godehusen, T. Richter, P. Wernet, and P. Zimmermann, "Open shells and multi-electron interactions: core level photoionization of the 3d metal atoms," J. Phys. B At. Mol. Opt. Phys. **39**, R79 (2006).
9. H. J. Eichler, P. Günter, and D. W. Pohl, *Laser-induced dynamic gratings*. (Berlin; New York: Springer-Verlag, 1986).
10. J. Salcedo, A. Siegman, D. Dlott, and M. Fayer, "Dynamics of Energy Transport in Molecular Crystals: The Picosecond Transient-Grating Method," Phys. Rev. Lett. **41**, 131 (1978).
11. R. J. D. Miller, R. Casalegno, K. A. Nelson, and M. D. Fayer, "Laser-induced ultrasonics: A dynamic holographic approach to the measurement of weak absorptions, optoelastic constants acoustic attenuation," Chem. Phys. **72**, 371-379 (1982).

12. T. F. Crimmins, A. A. Maznev, and K. A. Nelson, "Transient grating measurements of picosecond acoustic pulses in metal films," Appl. Phys. Lett. **74** 1344 (1999).
13. J. A. Rogers, A. A. Maznev, M. J. Banet, and K. A. Nelson, "Optical Generation and Characterization of Acoustic Waves in Thin Films: Fundamentals and Applications," Annu. Rev. Mater. Sci. **30**, 117 (2000).
14. G. D. Goodno, G. Dadusc, and R. J. D. Miller, "Ultrafast heterodyne-detected transient-grating spectroscopy using diffractive optics," J. Opt. Soc. Am. B **15**, 1791 (1998).
15. M. L. Cowan, J. P. Ogilvie, and R. J. D. Miller, "Two-dimensional spectroscopy using diffractive optics based phased-locked photon echoes," Chem. Phys. Lett. **386**, 184 (2004).
16. J. P. Ogilvie and K. J. Kubarych, "Chapter 5 Multidimensional Electronic and Vibrational Spectroscopy: An Ultrafast Probe of Molecular Relaxation and Reaction Dynamics," in *Advances In Atomic, Molecular, and Optical Physics*, vol. Volume 57, P. R. B. and C. C. L. E. Arimondo, Ed., pp. 249–321 (Academic Press, 2009).
17. R. I. Tobey, M. E. Siemens, M. M. Murnane, H. C. Kapteyn, D. H. Torchinsky, and K. A. Nelson, "Transient grating measurement of surface acoustic waves in thin metal films with extreme ultraviolet radiation," Appl. Phys. Lett. **89**, 091108 (2006).
18. R. I. Tobey, M. E. Siemens, O. Cohen, M. M. Murnane, H. C. Kapteyn, and K. A. Nelson, "Ultrafast extreme ultraviolet holography: dynamic monitoring of surface deformation," Opt. Lett. **32**, 286 (2007).
19. Q. Li, K. Hoogeboom-Pot, D. Nardi, M. M. Murnane, H. C. Kapteyn, M. E. Siemens, E. H. Anderson, O. Hellwig, E. Dobisz, B. Gurney, R. Yang, and K. A. Nelson, "Generation and control of ultrashort-wavelength two-dimensional surface acoustic waves at nanoscale interfaces," Phys. Rev. B **85**, 195431 (2012).
20. F. Morin, "Oxides Which Show a Metal-to-Insulator Transition at the Neel Temperature," Phys. Rev. Lett. **3**, 34 (1959).
21. Z. Yang, C. Ko, and S. Ramanathan, "Oxide Electronics Utilizing Ultrafast Metal-Insulator Transitions," Annu. Rev. Mater. Res. **41**, 337 (2011).
22. C. Kübler, H. Ehrke, R. Huber, R. Lopez, A. Halabica, R. F. Haglund, and A. Leitenstorfer, "Coherent Structural Dynamics and Electronic Correlations during an Ultrafast Insulator-to-Metal Phase Transition in VO2," Phys. Rev. Lett. **99**, 116401 (2007).
23. M. Imada, A. Fujimori, and Y. Tokura, "Metal-insulator transitions," Rev. Mod. Phys. **70**, 1039 (1998).
24. A. Cavalleri, T. Dekorsy, H. Chong, J. Kieffer, and R. Schoenlein, "Evidence for a structurally-driven insulator-to-metal transition in VO2: A view from the ultrafast timescale," Phys. Rev. B **70**, 161102 (2004).
25. A. Cavalleri, M. Rini, and R. W. Schoenlein, "Ultra-Broadband Femtosecond Measurements of the Photo-Induced Phase Transition in VO₂: From the Mid-IR to the Hard X-rays," J. Phys. Soc. Jpn. **75**, 011004 (2006).
26. N. B. Aetukuri, A. X. Gray, M. Drouard, M. Cossale, L. Gao, A. H. Reid, R. Kukreja, H. Ohldag, C. A. Jenkins, E. Arenholz, K. P. Roche, H. A. Dürr, M. G. Samant, and S. S. P. Parkin, "Control of the metal-insulator transition in vanadium dioxide by modifying orbital occupancy," Nat. Phys. **9**, 661 (2013).
27. A. Cavalleri, M. Rini, H. Chong, S. Fourmaux, T. Glover, P. Heimann, J. Kieffer, and R. Schoenlein, "Band-Selective Measurements of Electron Dynamics in VO2 Using Femtosecond Near-Edge X-Ray Absorption," Phys. Rev. Lett. **95**, 067405 (2005).
28. J. Jeong, N. Aetukuri, T. Graf, T. D. Schladt, M. G. Samant, and S. S. P. Parkin, "Suppression of Metal-Insulator Transition in VO2 by Electric Field-Induced Oxygen Vacancy Formation," Science **339**, 1402 (2013).
29. J. Grilj, E. Sistrunk, M. Koch, and M. Gühr, "A Beamline for Time-Resolved Extreme Ultraviolet and Soft X-Ray Spectroscopy," J. Anal. Bioanal. Tech. s12:005 (2014).
30. S. Shin, S. Suga, M. Taniguchi, M. Fujisawa, H. Kanzaki, A. Fujimori, H. Daimon, Y. Ueda, K. Kosuge, and S. Kachi, "Vacuum-ultraviolet reflectance and photoemission study of the metal-insulator phase transitions in VO2, V6O13, and V2O3," Phys. Rev. B **41**, 4993 (1990).
31. B. L. Henke, E. M. Gullikson, and J. C. Davis, "X-Ray Interactions: Photoabsorption, Scattering, Transmission, and Reflection at E = 50-30,000 eV, Z = 1-92," At. Data Nucl. Data Tables **54**, 181 (1993).
32. D. Kucharczyk and T. Niklewski, "Accurate X-ray determination of the lattice parameters and the thermal expansion coefficients of VO2 near the transition temperature," J. Appl. Crystallogr. **12**, 370 (1979).

33. C. N. Berglund and H. J. Guggenheim, "Electronic Properties of VO2 near the Semiconductor-Metal Transition," Phys. Rev. **185**, 1022 (1969).
34. C. Leroux, G. Nihoul, and G. Van Tendeloo, "From VO2(B) to VO2(R): Theoretical structures of VO2 polymorphs and in situ electron microscopy," Phys. Rev. B **57**, 5111 (1998).
35. J. W. Goodman, *Introduction to Fourier optics*, 3rd ed. (Englewood, CO: Roberts & Co, 2005).

Figures

Figure1

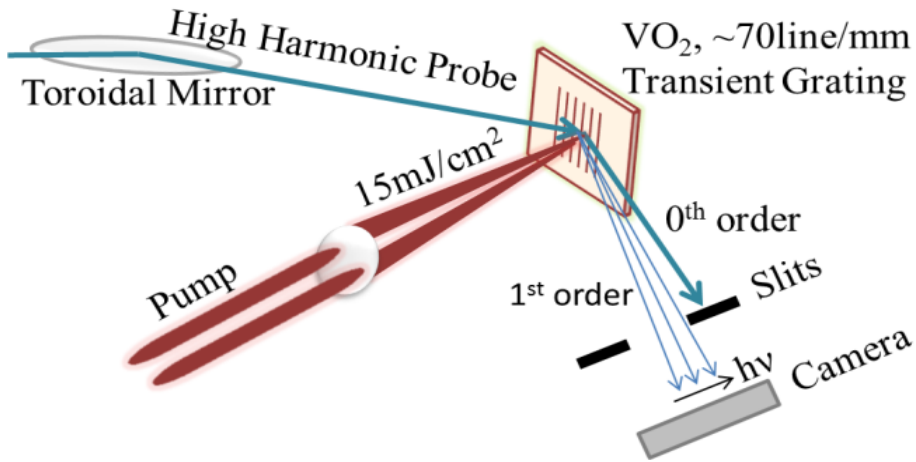

Figure 1: Transient grating experiment geometry. A transient grating with ~15µm period is created on the VO$_2$ surface by two 800nm beams. High harmonics generated in a high pressure gas cell are refocused using a toroidal mirror. First order diffraction from the sample surface is detected with a CCD camera.

Figure2

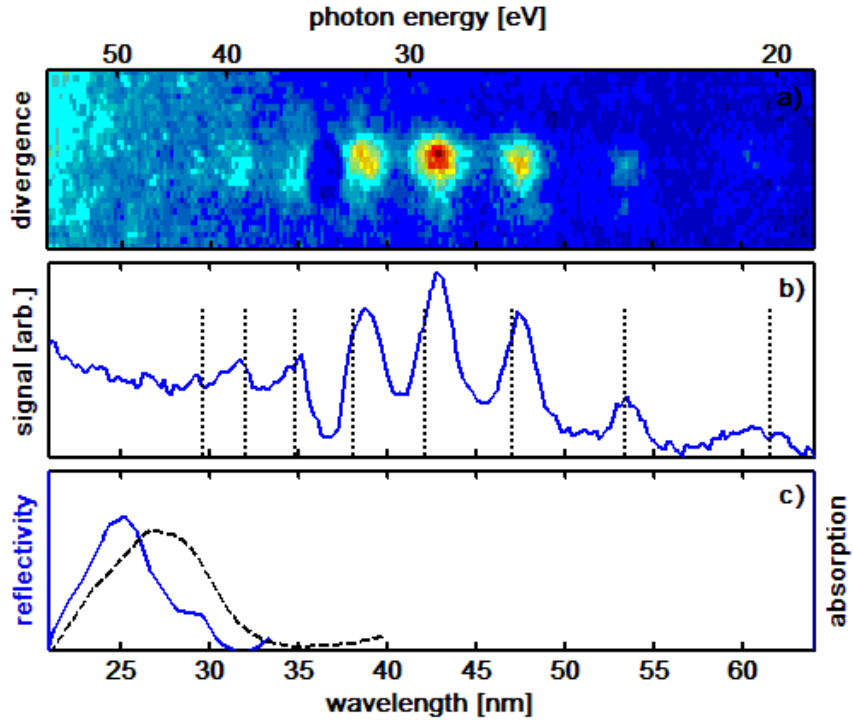

Figure 2: Typical first order diffraction signal from $VO_2$. a) Raw CCD image of 2.5s exposure on CCD camera. b) Integrated TG spectrum from image in a. Dashed lines indicate the calculated positions of harmonic orders 13-27. c) Reflectivity of $VO_2$ reproduced from Ref. [30] and absorption at V M-edge (dashed) calculated from values tabulated in Ref. [31].

Figure 3

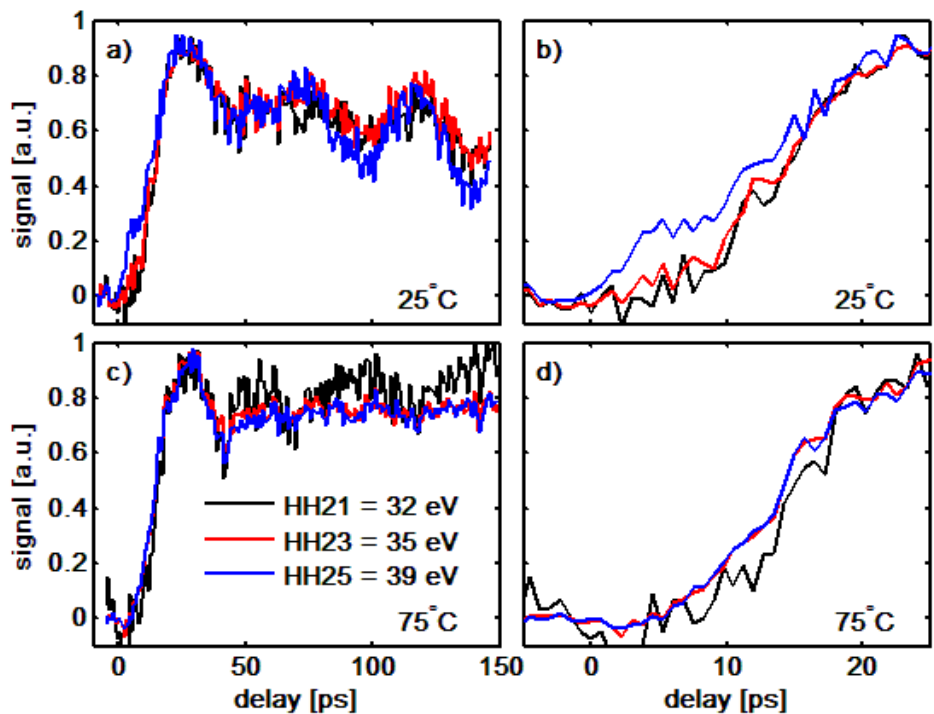

Figure 3: Time-integrated transient grating signal as a function of photon energy. The top (bottom) row shows spectrally integrated data for a sample below (above) the IMT temperature. b) and d) show close-ups of early time delays where a prompt response of HH25 is only present in the cold sample.